\documentclass[mathleft]{an}
\usepackage{graphicx}
\usepackage{times}
\usepackage{soul} 
\newcommand{\kms}{$\mathrm{\,km\,s^{-1}}$}

\begin{document}

\Pagespan{789}{}
\Yearpublication{2006}%
\Yearsubmission{2005}%
\Month{11}%
\Volume{999}%
\Issue{88}%

\title{Photometric and spectroscopic observations of three rapidly rotating 
late-type stars: EY~Dra, V374~Peg and GSC~02038-00293\,\thanks{Based on 
observations made with the Nordic Optical Telescope, La Palma, Spain; Isaac 
Newton Telescope of the Isaac Newton Group of Telescopes, La Palma, Spain; 
60-cm and 1-m telescopes of Konkoly Observatory, Hungary }}

\author{H. Korhonen\inst{1}\fnmsep\thanks{Corresponding author:
  \email{hkorhone@eso.org}}
\and K. Vida\inst{2}
\and M. Husarik\inst{3}
\and S. Mahajan\inst{4}
\and D. Szczygie{\l}\inst{5}
\and K. Ol{\'a}h\inst{2}}
\titlerunning{Photometric and spectroscopic observations of late type stars}
\authorrunning{Korhonen et al.}
\institute{
  European Southern Observatory, Karl-Schwarzschild Str. 2, D-85748 
  Garching bei Munchen, Germany
  \and 
  Konkoly Observatory of the Hungarian Academy of Sciences, H-1525 Budapest, 
  Hungary
  \and 
  Astronomical Institute of the Slovak Academy of Sciences, 059 60 
  Tatransk{\'a} Lomnica, The Slovak Republic
  \and
  School of Physics and Astronomy, University of Birmingham, Edgbaston, 
  Birmingham B15 2TT, United Kingdom
  \and
  Department of Astronomy, The Ohio State University, 140 W. 18th Ave., 
  Columbus OH 43210, USA
}

\received{-}
\accepted{-}
\publonline{-}


\abstract{Here, $BV(RI)_{C}$ broad band photometry and intermediate resolution 
spectroscopy in H$\alpha$ region are presented for two rapidly rotating 
late-type stars: EY~Dra and V374~Peg. For a third rapid rotator, 
GSC~02038-00293, intermediate resolution H$\alpha$ spectroscopy and low 
resolution spectroscopy are used for spectral classification and stellar 
parameter investigation of this poorly known object. The low resolution 
spectrum of GSC~02038-00293 clearly indicates that it is a K-type star. Its 
intermediate resolution spectrum can be best fitted with a model with 
T$_{\rm eff}$=4750~K and $v\,\sin i$=90\kms\ , indicating a very rapidly 
rotating mid-K star. The H$\alpha$ line strength is variable, indicating 
changing chromospheric emission on GSC~02038-00293. In the case of EY~Dra and 
V374~Peg, the stellar activity in the photosphere is investigated from the 
photometric observations, and in the chromosphere from the H$\alpha$ line. The 
enhanced chromospheric emission in EY~Dra correlates well with the location of 
the photospheric active regions, indicating that these features are spatially 
collocated. Hints of this behaviour are also seen in V374~Peg, but it cannot 
be confirmed from the current data. The photospheric activity patterns in 
EY~Dra are stable during one observing run lasting several nights, whereas in 
V374~Peg large night-to-night variations are seen. Two large flares, one in the
H$\alpha$ observations and one from the broadband photometry, and twelve 
smaller ones were detected in V374~Peg during the observations spanning nine 
nights. The energy of the photometrically detected largest flare is estimated 
to be $4.25\times 10^{31}-4.3\times 10^{32}$ ergs, depending on the waveband. 
Comparing the activity patterns in these two stars, which are just below and 
above the mass limit of full convection, is crucial for understanding dynamo 
operation in stars with different internal structures.
}

\maketitle

\section{Introduction}

Among low-mass stars, the limit of full convection plays an important role. On 
the two sides of this limit stars are thought to host different kind of 
dynamos: according to theories, stars above $\sim$0.35\,M$_\odot$ have 
radiative core and a convective envelope (Chabrier~\& Baraffe \cite{cha_bar}), 
and create their magnetic field with an $\alpha\Omega$-dynamo (e.g., Parker 
\cite{parker}; Babcock \cite{bab}; Leighton \cite{lei}), similarly to the 
Sun. Stars below this limit are fully convective, and are found to be rotating 
almost as rigid bodies (Barnes et al. \cite{barnes05}).

Theoretical studies of K{\"u}ker \& R{\"u}diger (\cite{kue_rue}) and Cha\-brier
\& K{\"u}ker (\cite{cha_kue}) show, that fully convective stars rotating as 
solid bodies can produce large-scale, non-axi\-sym\-met\-ric fields using 
$\alpha^2$-dynamo. But Dobler, Stix \& Brandenburg (\cite{dobler06}) suggest 
that, these stars can develop axisymmetric poloidal magnetic fields, given 
that they have strong differential rotation. Regardless of their interior 
structure, both types show signs of activity (see, e.g., Delfosse et al. 
\cite{del_etal98}). 

In this paper we present observations of three rapidly rotating stars: EY~Dra, 
V374~Peg and GSC~02038-00293. It is especially intriguing to compare the 
observations of fully convective V374~Peg and EY~Dra which has a small 
radiative core. 

V374 Peg (RA 22 01 13.11, dec +28 18 24.9, V=11.89 mag) is a fully convective 
M4 dwarf with a mass of 0.28~M$_{\odot}$ (Delfosse et al.~\cite{del_etal00}), 
whose observed properties can question the existing dynamo theories. The 
unusual activity of the star has been observed by Greimel \& Robb 
(\cite{gre_rob}) and Batyrshinova \& Ibragimov (\cite{bat_ibr}), who reported 
frequent and intense flares on the object. Using spectropolarimetric methods 
Donati et al. (\cite{donati06}) and Morin et al. (\cite{morin08}) showed that 
V374 Peg is rotating almost as a rigid body, but at the same time has a stable,
axisymmetric poloi\-dal magnetic field contradicting the theoretical models.

EY Dra (RA 18 16 16.05, dec +54 10 16.0, V=11.83 mag) is a well-studied rapidly
rotating dM1-2e star (Jeffries, James, \& Bromage \cite{jef94}). Its mass, 
$\sim$0.49~M$_{\odot}$ (according to Eibe \cite{eibe}), implies that the star 
probably has a radiative core and convective outer en\-ve\-lope. Eibe 
(\cite{eibe}) published H$\alpha$ observations, and explained them with plages 
and co-rotating prominence clouds above the surface. Using Doppler images, 
Barnes \& Collier Cameron (\cite{bar_col}) showed that EY Dra has spots at all 
latitudes. They also detected differential rotation with a shear of 
$\Delta\Omega=0.0608$, assuming a solar-type differential rotation: 
$\Omega(\theta)=\Omega_0-\Delta\Omega\sin^2(\theta)$. Korhonen et al. 
(\cite{Kor_EYDra}) presented photometric observations in $V$ and $R$ filters, 
together with optical and near-infrared spectroscopy, and showed a possible 
association of photospheric starspots and chromospheric plages. Vida 
(\cite{vida07}) and Vida et al. (\cite{vida10}) showed evidence for spot 
evolution on the stellar surface using $\sim$1000-day-long photometric data, 
and an activity cycle with a period of $\sim$350 days. Signs of a possible 
flip-flop mechanism is also found, which might give indication of the type of 
the underlying dynamo mechanism producing the observed stellar activity (see, 
e.g., Elstner~\& Korhonen \cite{els_kor}).

GSC~02038-00293 (RA 16 02 48.229, dec +25 20 38.20, V=10.62 mag) is a very 
little studied star that was originally discovered during a programme 
identifying optical counter parts of ROSAT X-ray sources by Bernhard~\& Frank 
(\cite{ber_fra}). In the same paper photometric observations for 2005--2006 
were presented together with older ROTSE data. According to these observations
GSC~02038-00293 is an eclipsing RS~CVn type binary with the orbital 
period of 0.495410 days. This value is similar to the one given by the 
SuperWASP observations (Norton et al.~\cite{WASP}), and is confirmed by 
the 2007 data (Frank \& Bernhard \cite{fra_ber}). These investigations also
show that the star\-spot locations are stable over long time periods, but that 
the exact shape of the light-curve changes on a time scale of weeks. Also, 
hints of a 6--8 year long activity cycle are seen (Bernhard \& Frank 
\cite{ber_fra}). The spectral type of GSC~02038-00293 is determined to be K 
from low resolution spectra (Dragomir, Roy \& Rutledge \cite{drr}).

In this paper spectroscopic and photometric observations of these three stars 
are presented. For GSC~02038-00293 the low and intermediate resolution spectra 
are used to study the stellar parameters of this poorly known system. For 
EY~Dra and V374~Peg the photospheric activity patterns are deduced from 
broadband photometry and the chromospheric activity is studied using 
intermediate resolution observations of H$\alpha$ line. The activity seen at 
these different levels of the stellar atmosphere is correlated, and the 
activity patterns in fully convective V374~Peg and EY~Dra, which still has a 
small radiative core, are compared.

\section{Observations}

Medium resolution spectroscopy in the H$\alpha$ region was obtained at the 
2.6-m Nordic Optical Telescope (NOT) using Andalucia Faint Object Spectrograph 
and Camera (ALFOSC) and the Intermediate Dispersion Spectrograph (IDS) at the 
2.5-m Isaac Newton Telescope (INT) of the Isaac Newton Group of Telescopes. 
Both the telescopes are located in the Observatorio del Roque de los Muchachos 
of the Instituto de Astrofisica de Canarias, on La Palma, Spain. Broadband 
photometry of EY~Dra and V374~Peg was obtained with the 1-m RCC telescope in 
Piszk{\'e}stet{\H o} mountain station of the Hungarian Konkoly Observatory and 
the 0.6-m telescope of the Konkoly Observatory at Sv\'abhegy, Budapest. All the
observations of EY~Dra were phased using the ephemeris: 
\begin{equation}
HJD = 2453588.16582 + 0\fd 4587 \times E
\label{E_EYDra}
\end{equation} 
(used by Vida et al. \cite{vida10}), the V374~Peg observations with 
\begin{equation}
HJD = 2453601.78613 + 0\fd 4456 \times E
\label{E_V374Peg}
\end{equation}
(from Morin et al. \cite{morin08}), and the GSC~02038-00293 data using
\begin{equation}
HJD = 2453560.491 + 0\fd 495410 \times E
\label{E_GSC}
\end{equation}
(from Bernhard \& Frank \cite{ber_fra}; Frank \& Bernhard \cite{fra_ber}).

\subsection{Spectroscopy}

The ALFOSC observations of EY~Dra and V374~Peg were obtained using grism\#17, 
and a 0.5 arcsec slit during the nights starting 2008 June 24 and June 28. 
With this instrument configuration a resolving power $\lambda/\Delta\lambda$ 
of 10,000 and the wavelength coverage of 6335--6860~{\AA} is obtained. The 
detector which is an E2V Technologies 2k back-illuminated CCD with 13.5$\mu$m 
pixels shows a strong fringing pattern in the spectrum, which is visible 
redwards of $\sim$6400~{\AA}. To remove the fringing pattern, observations were
made in sets of five exposures with moving the target along the slit between 
consecutive exposures. After every five object exposures, a Halogen flat field 
and a Neon arc spectrum were obtained. In the analysis only the combined 
spectra are used.

During the night starting 2008 June 24, observations of GSC~02038-00293 were 
also obtained with ALFOSC using the same setup and observing strategy as for 
EY~Dra and V374~Peg, and during the night starting 2008 June 28 using 
grism\#4 and a 1.3 arcsec slit. The latter instrument configuration gives 
spectral coverage 3200--9100~{\AA} and a resolving power of $\sim$700, but with
a fringing pattern that gets progressively worse redwards of 6700~{\AA}. 
Observations of the spectrophotometric standard HD~338808 were also obtained 
with the same instrument settings. A detailed observing log for the 
intermediate resolution observations of GSC~02038-0029 is given in 
Table~\ref{GSC}.

\begin{table}
\caption{Details of the intermediate resolution spectroscopic observations of 
  GSC~02038-00293 obtained on the night starting 2008 June 24. The UT time at 
  the mid point of the observations, heliocentric Julian date, rotational 
  phase using Eq.~\ref{E_GSC}, and the number of observations combined 
  together for each phase. The exposure time for each individual observation 
  is 20 seconds, thus giving total exposure time for each phase of 100 or 80 
  seconds.} 
\label{GSC}
\begin{center}     
\begin{tabular}{l l l l}
\hline
  UT       & HJD       & phase  & No of \\
  midpoint & 2454600+  &        & spectra  \\
\hline
21:51:58   & 42.409550 & 0.8852 & 5 \\
22:18:16   & 42.427811 & 0.9220 & 5 \\
22:42:29   & 42.444633 & 0.9560 & 5 \\ 
23:39:42   & 42.484361 & 0.0362 & 5 \\
23:52:01   & 42.492924 & 0.0535 & 4 \\ 
\hline
\end{tabular}
\end{center}
\end{table}

The IDS on INT was only used to observe EY~Dra and V374~Peg during the night 
starting 2008 June 26. Grating H1800V centred at 6600~{\AA} was used together 
with a slit width of 1.4 arcsec, this combination gives spectral coverage of 
6280--6962~{\AA} and resolving power of 10,000. 

All the observations were reduced using Image Reduction and Analysis Facility 
(IRAF) which is distributed by KPNO/NOAO.

\begin{figure}
\includegraphics[width=80mm]{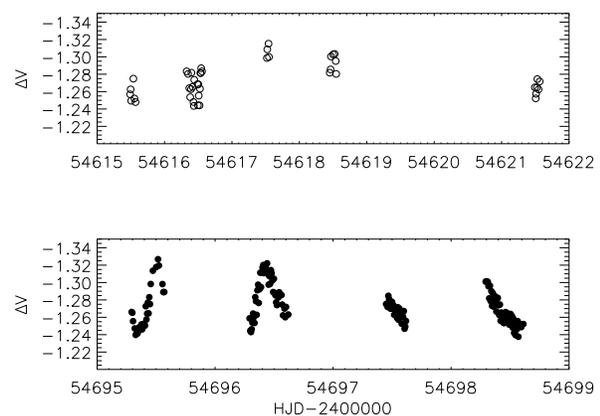}
\caption{The EY Dra $V$ magnitudes plotted against Julian date.}
\label{EYDra_JD}
\end{figure}

\begin{figure*}
\includegraphics[height=170mm, angle=270]{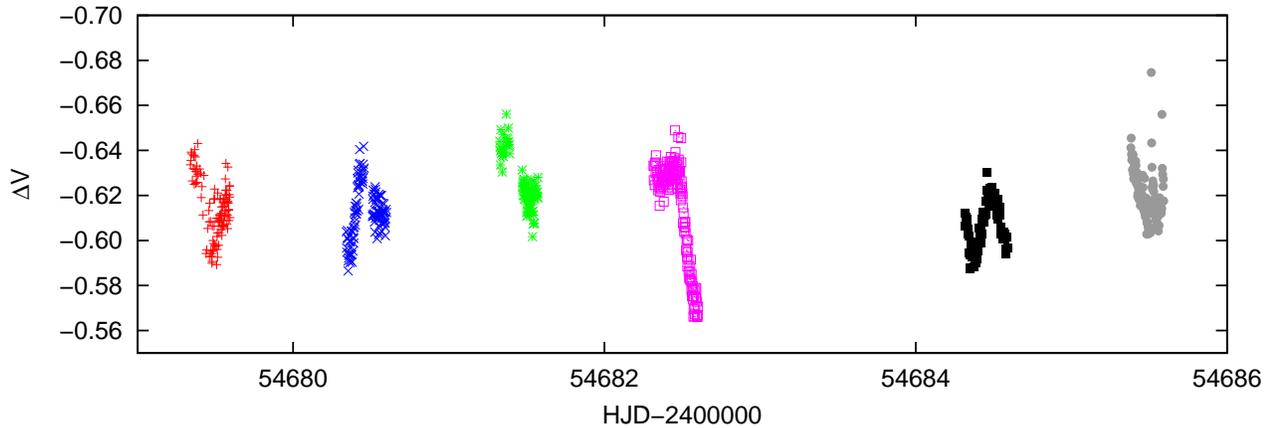}
\caption{The V374~Peg $V$ magnitudes plotted against Julian date.}
\label{V374Peg_JD}
\end{figure*}

\subsection{Photometry}

In addition to the spectroscopy photometric observations in $B$, $V$, $R_{C}$ 
and $I_{C}$-bands are presented for EY~Dra (published also in Vida et 
al.~\cite{vida10}) and V374~Peg. The photometry of EY~Dra was obtained with 
the 0.6-m telescope. The telescope is equipped with a Wright Instruments 
$750 \times 1100$ CCD. The $V$-band instrumental magnitudes plotted against 
Julian date in Fig.\,\ref{EYDra_JD} show measurements from two observing runs 
before and after the spectroscopic observations: five nights between JDs 
2454615--2454622 (2008 May 29--June 3) and four between JDs 2454695--2454699 
(2008 August 16--20). For observing V374~Peg the 1-m RCC telescope with a 
Princeton $1300\times 1300$ CCD was used. The light-curve in 
Fig.\,\ref{V374Peg_JD} was obtained from observations taken over six nights 
between JDs 2454679--2454686 (2008 July 21--August 7).

\section{Results}

\subsection{Photospheric spots from the broad band photometry}

\begin{figure}
\includegraphics[width=80mm]{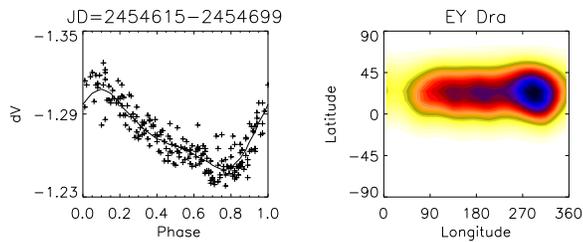}
\caption{Phased light-curve of EY Dra. Left: $V$ magnitudes plotted against 
  the phase. Right: spot-filling factor map from the light-curve inversion. 
  Darker colour indicates a larger filling-factor value.}
\label{EYDra_V}
\end{figure}

As is seen in the Fig.~\ref{EYDra_JD}, the photometry of EY~Dra shows clear 
variability. If these observations are phased using the ephemeris given in 
Eq.~\ref{E_EYDra} a broad minimum around phases 0.4--0.9 is seen 
(Fig.~\ref{EYDra_V}, left panel). 

To investigate the longitudinal spot\footnote{Note that throughout the paper 
term spot does not strictly mean one sunspot-like structure, but can also be an
active region consisting of several individual spots. There is no way to 
distinguish between these cases from light-curves} configuation in detail 
light-curve inversion techniques (see, e.g., Ol{\'a}h et al. \cite{olah06}) 
were used to produce a map of the stellar surface showing the fraction of each 
pixel covered by spots. This so-called spot-filling factor map shows a 
detailed longitudinal spot distribution, and in some cases, reveals the 
existence of close-by spots that would not otherwise be separable (Savanov 
\& Strassmeier \cite{sav_str}). But due to the one dimensional nature of the 
light-curve, no latitudinal spot information can be obtained from these maps.
The resulting spot-filling factor map for EY~Dra can be seen in the right panel
of Fig.~\ref{EYDra_V}. The following parameters were used in the 
inversion: inclination 66$^{\circ}$ (Robb \& Cardinal \cite{robb_card}), 
unspotted surface temperature 4000~K (adopted from 3900~K determined by Barnes 
\& Collier Cameron \cite{bar_col}), temperature of spots 3000~K (based on the 
typical difference between the spot and unspotted surface temperatures in 
active stars) and limb-darkening coefficient of 0.81 (Al-Naimyi \cite{aln}).
The map shows primary spot at phase 0.8 (longitude 290$^{\circ}$) and an 
extended secondary spot structure concentrated on phase 0.48 (longitude 
170$^{\circ}$). This result is similar to the one shown in Korhonen et al. 
(\cite{Kor_EYDra}) and Vida et al. (\cite{vida10}). In both cases two active 
regions on the surface were seen with separation of 0.3--0.5 in phase 
(110--180$^{\circ}$). 

\begin{figure}
\includegraphics[width=80mm]{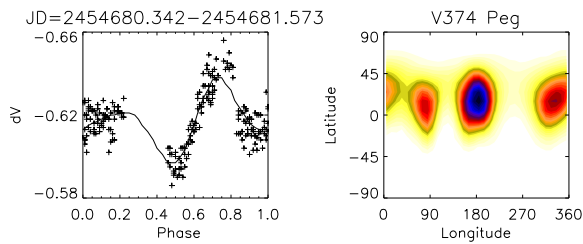}
\caption{The same as Fig.~\ref{EYDra_V}, but now for V374~Peg. Due to the fast
  variability in the light-curve only data from two nights have been used 
  (nights starting 2008 August 1 and August 2).}
\label{V374Peg_V}
\end{figure}

The $V$-band light-curve of V374~Peg for two consecutive nights is presented in
the left panel of Fig.~\ref{V374Peg_V}. The data were phased using ephemeris 
from Eq.~\ref{E_V374Peg}. The light-curve minimum occurs at phases 0.3--0.6 
(longitude 110--215$^{\circ}$). The exact shape of the minimum changes rapidly 
from night-to-night (also see discussion in Section~\ref{chromo}). Due to this 
fast variability, and incomplete phase coverage during a single night, 
spot-filling factor map for V374~Peg is only obtained from observations 
spanning two consecutive nights during which the behaviour of the light-curve 
is similar and stable (JD~2454680.342--2454681.573). Unfortunately this still 
leaves a gap of 0.2 in phase centred around the phase 0.3 (longitude 
110$^{\circ}$). The parameters adopted for the light-curve inversion are 
the same as for EY~Dra, except the inclination which is 70$^{\circ}$ (Donati et
al. \cite{donati06}).  Due to the relatively large gap the solution of the 
light-curve inversion is not stable around these phases. But even then, it is 
clear from the spot-filling factor map presented in the right panel of 
Fig.~\ref{V374Peg_V}, that the spots on V374~Peg concentrate at the phase range
0.9--0.6 (longitude 320--215$^{\circ}$), and that the main spot is centred at 
phase 0.50 (longitude 180$^{\circ}$). A secondary spot is also clearly seen 
centred at phase 0.94 (longitude 340$^{\circ}$), but the shape and extent of 
the other secondary spot seen centred at phase 0.21 (longitude 75$^{\circ}$) 
cannot be determined accurately due to the missing data. It is also plausible 
that the two secondary spots form one large active region, and not two separate
 ones.

\begin{figure}
\includegraphics[width=70mm]{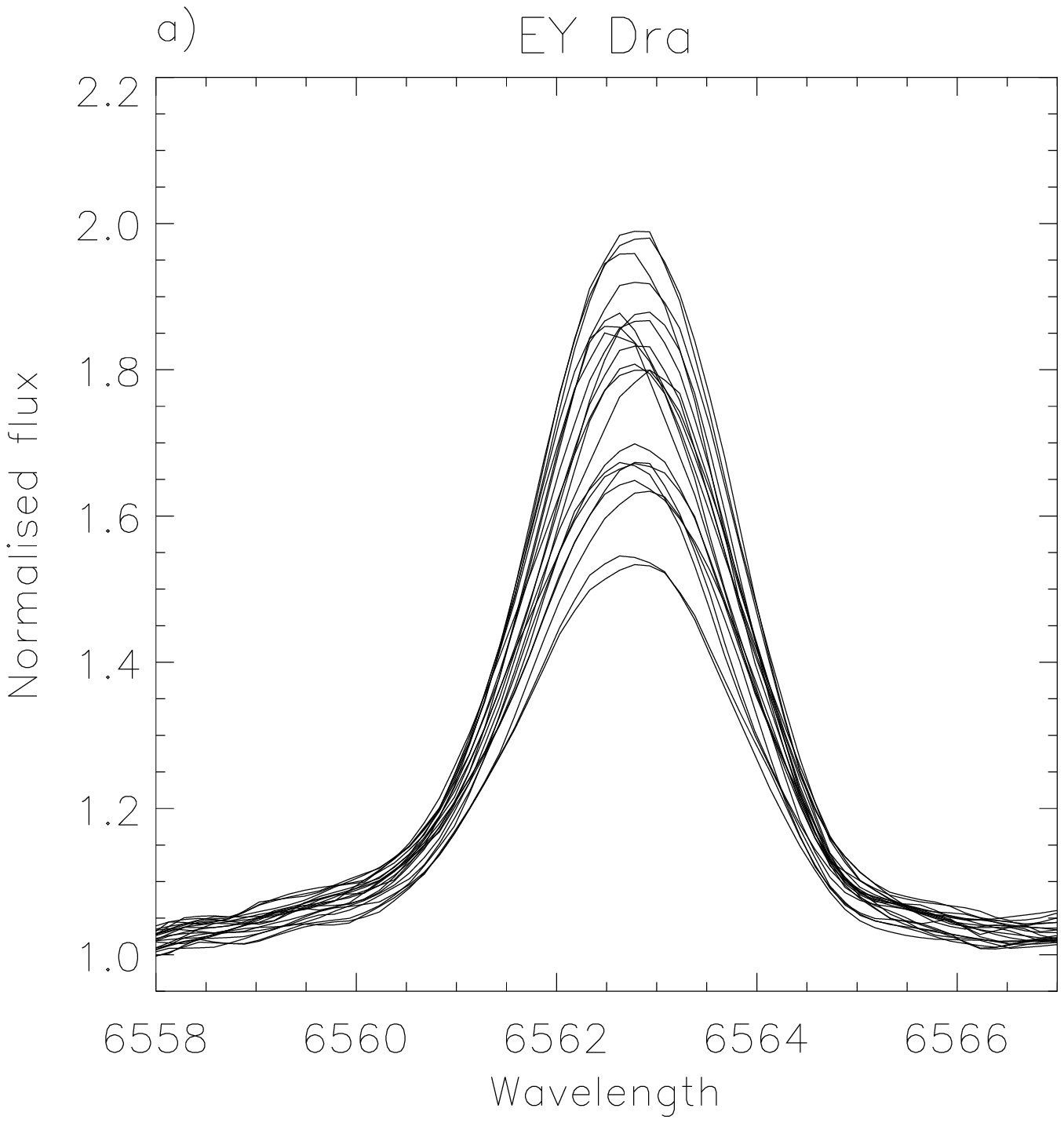}
\includegraphics[width=70mm]{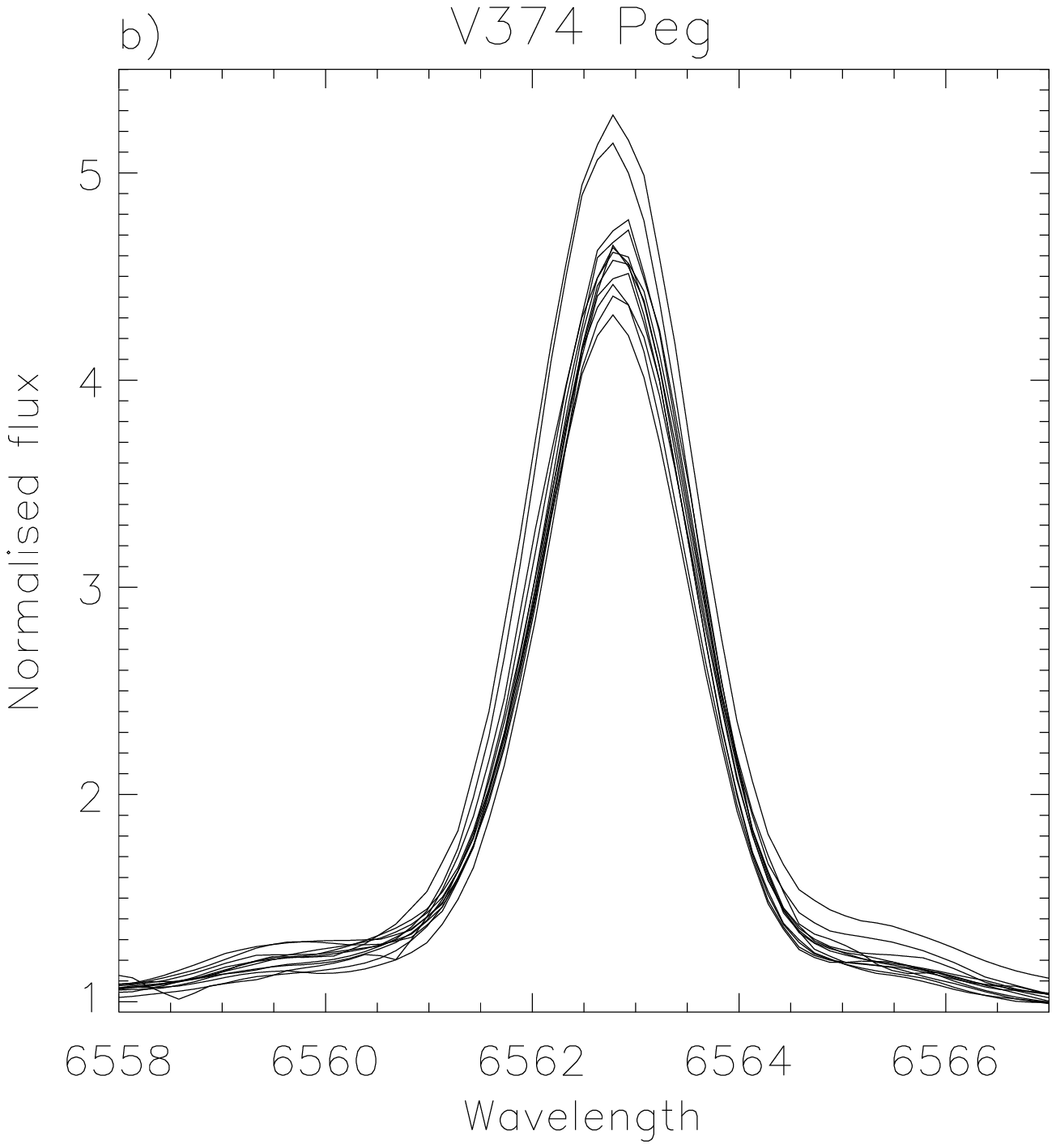}
\caption{The H$\alpha$ line profiles. a) All the H$\alpha$ lines of EY~Dra 
plotted into the same figure to show the changes in the strength of the line. 
b) The same as a) but for V374~Peg. Note that the flux scales are different.}
\label{both_Ha}
\end{figure}

\subsection{H$\alpha$ line variability}

In EY~Dra the H$\alpha$ line is an emission line that shows clear variability, 
as is also shown in the Fig.~\ref{both_Ha}. The earlier studies have shown 
prominences and cool clouds in the chromosphere (see, e.g., Eibe \cite{eibe}). 
Also, in V374~Peg the H$\alpha$ line is in emission and variable, as shown in 
the right panel of Fig.~\ref{both_Ha}. The H$\alpha$ variability is 
investigated in detail in Section~\ref{chromo} and compared to the 
behaviour of the photospheric spots.

\begin{figure}
\includegraphics[width=80mm]{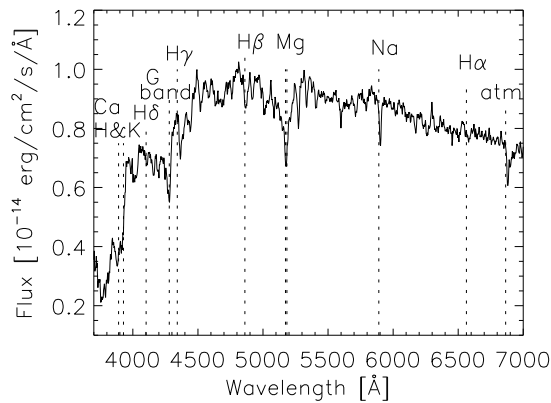}
\caption{The low resolution spectrum of GSC~02038-00293 in flux scale. The main
  spectral features have been identified.}
\label{GSC_flux}
\end{figure}

\begin{figure*}
\includegraphics[width=160mm]{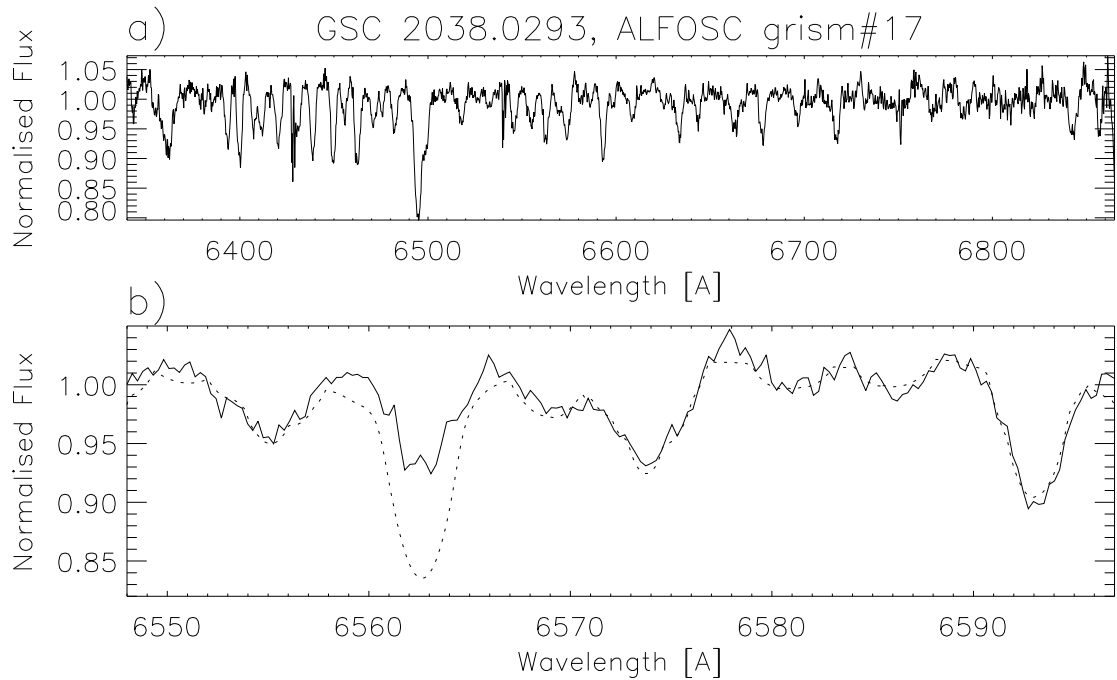}
\caption{The intermediate resolution spectrum of GSC~02038-00293. a) The whole 
  spectrum obtained with grism\#17. b) Smaller region of the same spectrum 
  (solid line) together with a model (dotted line) calculated using 
  T$_{\rm eff}$=4750~K, $\log g$=4.5, microturbulence velocity 1.5\kms, and 
  v$\sin i$=90\kms.}
\label{GSC_Ha}
\end{figure*}

\subsection{GSC~02038-00293}

The low resolution spectrum of GSC~02038-00293 which is shown in 
Fig.~\ref{GSC_flux} was used for spectral classification of this target. The 
main spectral features seen are neutral metals, like Mg and Na. Weak Balmer 
lines are also present, but no clear molecular bands are detected. All these 
features indicate a K-type star, as was also deduced by Dragomir, Roy \& 
Rutledge~(\cite{drr}) from earlier low resolution spectra.

The intermediate resolution spectrum of GSC~02038-00293 in the H$\alpha$ 
region, after combining the five individual observations, is shown in 
Fig.~\ref{GSC_Ha}a. As can be seen, the spectral lines are very broad and 
H$\alpha$ itself is a very weak absorption line. Synthetic stellar spectra were
calculated using SPECTRUM code (Gray \& Corbally \cite{SPECTRUM}) and Kurucz 
model atmospheres (Kurucz \cite{kurucz}). For the calculations $\log g =4.5$ 
was adopted. The best fit to the observed spectrum was obtained for 
T$_{\rm eff}=4750 \pm 250$~K, microturbulence velocity $\xi=1.5$\kms, and 
$v\,\sin i = 90 \pm 10$\kms (see Fig.~\ref{GSC_Ha}b). Due to the low 
resolution and poor signal-to-noise ratio, the errors in these parameters are 
quite large. Using the $v\,\sin i$ measured in this work and the rotation 
period of 0.495410 days (Bernhard \& Frank \cite{ber_fra}), the estimated 
radius of GSC~02038-00293 is 
${\rm R}\times\sin i = 0.88 \pm 0.10 {\rm R}_{\odot}$, 
implying spectral type of late-G or later. 

\begin{figure}
\includegraphics[width=80mm]{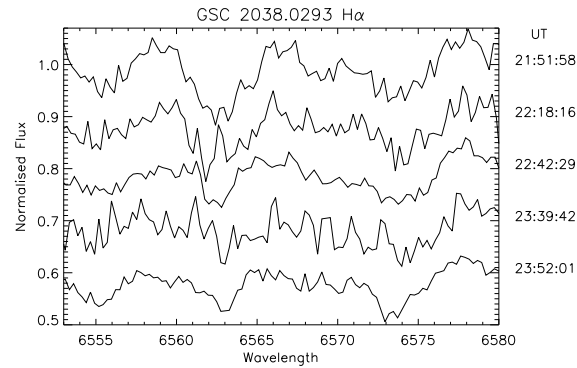}
\caption{The five separate exposures of GSC~02038-00293 in the H$\alpha$ 
  region, from the night starting 2008 June 24. The mid point of the 
  observations are given on the right side of the plot. The spectra are offset 
  from each other by 0.1 in normalised flux.}
\label{GSC_variab}
\end{figure}

The comparison of the averaged H$\alpha$ line spectrum to the model shows that 
this line in GSC~02038-00293 is clearly weaker than expected at this 
temperature and $\log g$. This could be explained by magnetic activity in which
the chromospheric activity is partly filling in the line core. The five 
individual exposures of the GSC~02038-00293 H$\alpha$ region show clear time
variability, as seen from Fig.~\ref{GSC_variab}. The first exposure shows clear
and strong absorption line. With time, the line gets progressively shallower, 
thus explaining the weak H$\alpha$ line seen when combining all the 
observations to a better signal-to-noise spectrum (shown in Fig.~\ref{GSC_Ha}).
This implies variable chromospheric emission in GSC~02038-00293. The hypothesis
is further strengthened by the reported prominent Ca H\&K emission cores 
observed in GSC 02038-00293 (Dragomir, Roy~\& Rutledge~\cite{drr})

It is worth noting that the line-profiles of GSC~02038-00293 are slightly 
triangular. This could imply a secondary component in this system which may 
affect the spectral line profiles. Since this system is thought to be a 
RS~CVn-type binary (Bernhard \& Frank \cite{ber_fra}; Frank \& Bernhard 
\cite{fra_ber}; Eker et al. \cite{CABS08}), it is not unlikely that there 
is a spectroscopically detectable companion. But, observations at other 
epochs are needed to confirm this hypothesis. Also, if the system is a 
spectroscopic binary, and the secondary during these observations was in such 
an orbital phase that its spectral lines are superimposed with the primary 
lines, the stellar parameters determined here (effective temperature and 
v$\sin i$) would also be affected by the secondary.

\section{Discussion}

\subsection{Flares}

\begin{figure}
\includegraphics[width=70mm]{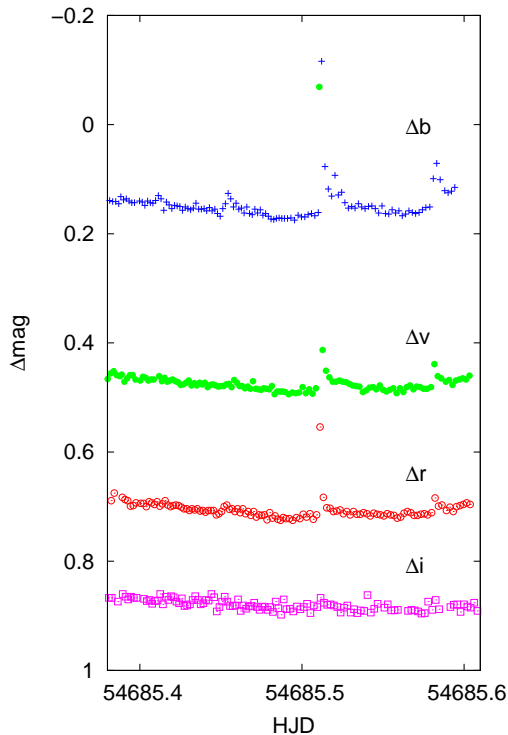}
\caption{Flares seen in V374~Peg from the photometric observations of the 
night starting 2008 August 6. The light-curves have been shifted arbitrarily. 
The light-curves show instrumental values (see text).}
\label{photom_flare}
\end{figure}

On JD2454685 (2008 August 6) an energetic flare was observed on V374 Peg 
(see Fig.~\ref{photom_flare}). This event lasted $\sim$1~hour, and could be 
observed in all the filters, $BV(RI)_C$. The peak of the flare happened after a
$B$ exposure, so in this filter only the declining phase was detected. For 
estimating the energy of the flare the method described by K{\H o}v{\'a}ri et 
al. (\cite{kovari07}) was applied. Instrumental magnitudes were used for 
calculations, since the interpolation needed for transforming to the 
international system would blur the fast changes in the light-curve. The 
resulting energies are $2.04\times 10^{32}$, $3.00\times 10^{32}$, $4.30\times 
10^{32}$, and $4.25\times 10^{31}$ergs in $B$, $V$, $R_C$, and $I_C$ filters, 
respectively. As can be seen in Fig.~\ref{photom_flare}, additional smaller 
flares occur before and after this major event. 

Another flare in V374~Peg is detected in H$\alpha$ around phase 0.2 in the 
observations from the night starting 2008 June 26. This flare can be seen in 
the equivalent width measurements shown in Fig.~\ref{V374Peg_V_Ha}b. 
Unfortunately, only the end stages of the flare are recorded. Thus, no 
information on the strength and the length of the flare can be given.

\begin{figure}
\includegraphics[width=70mm]{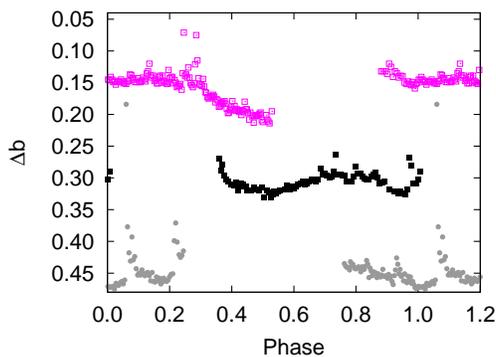}
\caption{The phased $B$ light-curves from Julian dates 2454682--2454685. The 
light-curves from different nights have been shifted, to show better the 
behaviour. Twelve small flares are seen in these data around phases 0.9--0.3.
The plot shows light-curves with instrumental magnitudes. }
\label{flare2}
\end{figure}

Besides these two energetic flares, twelve smaller events can be seen in 
the photometric data on the last three nights. In Fig.~\ref{flare2} the flares 
in $B$ band light-curves for the last three observing nights are shown. Most 
flares seem to erupt in the phases 0.9--0.3. The observed H$\alpha$ flare also
occurs at the same phase range. Though, smaller events can happen at all 
phases. The night-to-night light-curve changes and the continuous flaring are 
clearly related showing the very active nature of this star.

\subsection{Correlating photospheric and chromospheric activity}
\label{chromo}

The photometric observations of EY~Dra are not taken at the same time as the 
spectroscopic ones, but the observations obtained $\sim$25 days before the 
spectroscopy show the same light-curve shape as the photometry obtained 
$\sim$55 days after the spectroscopic observations. Therefore, the observations
from different epochs can be compared.

\begin{figure}
\includegraphics[width=70mm]{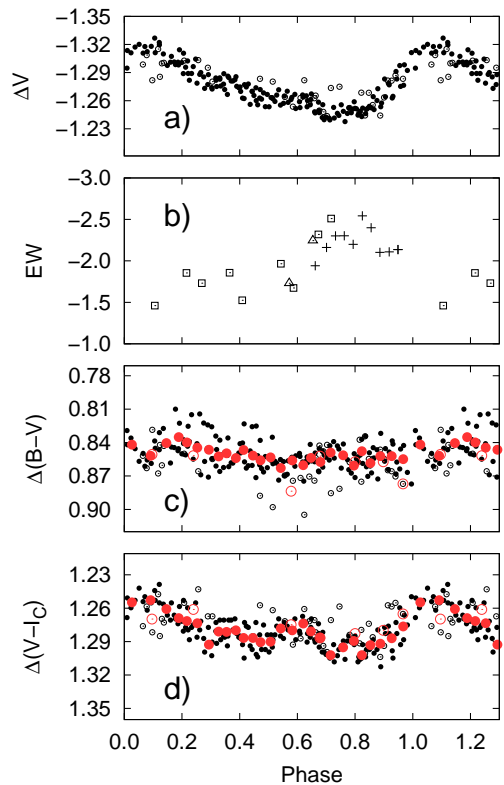}
\caption{The comparison of the photospheric and chromospheric activity in 
EY~Dra. a) The $V$ magnitudes plotted against the rotational phase. Filled and 
empty circles show observations from different epochs (see 
Fig.~\ref{EYDra_JD}.) b) The H$\alpha$ equivalent widths plotted against 
rotational phase. Different symbols denote observations from different nights. 
Triangles and boxes show NOT observations from June 24 and June 28, 
respectively, crosses mark the INT data from June 26. c) The $B$-$V$ colours 
plotted against the phase. The larger symbols are an average of typically five 
individual observations obtained close in time. d) Same as c) but for $V$-$I$.}
\label{EYDra_V_Ha}
\end{figure}

The phased light-curve of EY~Dra is shown in Fig.~\ref{EYDra_V_Ha}a together 
with the equivalent widths measured from the H$\alpha$ line profiles 
(Fig.~\ref{EYDra_V_Ha}b), and the phased colour index curves of EY~Dra in 
$B$-$V$ (Fig.~\ref{EYDra_V_Ha}c) and $V$-$I$ (Fig.~\ref{EYDra_V_Ha}d). The 
$B$-$V$ is practically flat except for a small increase at phases 0.1--0.3.
Just like the $V$-band light-curve, $V$-$I$ curve also indicates two cool
spots around 0.4 and 0.8 phases (see Fig.~\ref{EYDra_V}).

The comparison clearly shows that at the phases of the photospheric spots, more
emission is seen in the chromosphere. This configuration implies that the 
chromospheric plages on EY~Dra concentrate at the locations of photospheric 
spots, as is also the case for Sun. It is only recently that this behaviour
has been observed in several active stars, e.g., RS~CVn-type binaries 
$\lambda$~And and II~Peg (Frasca et al.~\cite{frasca_bin}), rapidly-rotating 
K1-dwarf LQ~Hya (Frasca et al.~\cite{frasca_LQHya}), T~Tauri-type star TWA~6 
(Skelly et al.~\cite{skelly}) and for a young, late-type star, SAO~51891 
(Biazzo et al.~\cite{biazzo09}).

Also, for V374~Peg the spectroscopic and photometric observations are not 
contemporaneous, but the spectra have been obtained $\sim$40 days before the 
photometric observations. It has been shown elsewhere (e.g., Morin et al. 
\cite{morin08}) that the active regions in V374 Peg do not change quickly. 
Hence, the photometry and spectroscopy can be compared.

\begin{figure}
\includegraphics[width=70mm]{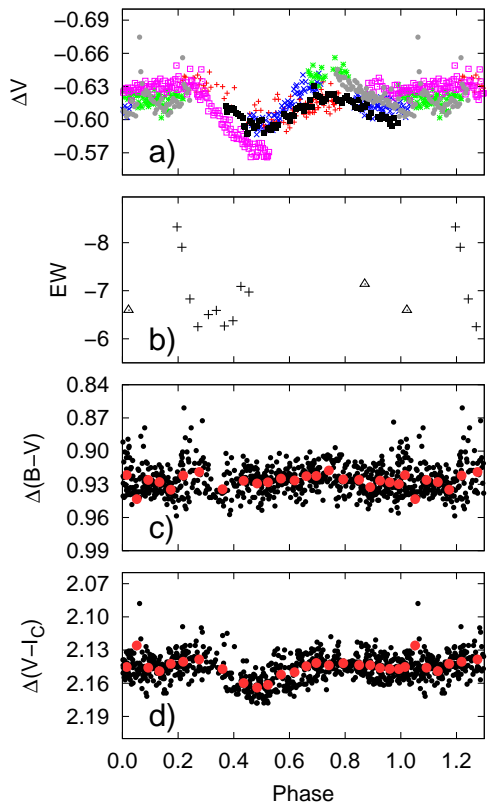}
\caption{The same as Fig.~\ref{EYDra_V_Ha}, but now for V374~Peg. Panel a) show
  different nights with different symbols and colours (see 
  Fig.~\ref{V374Peg_JD}).}
\label{V374Peg_V_Ha}
\end{figure}

In Fig.~\ref{V374Peg_V_Ha} the $V$-band light-curve (Fig.~\ref{V374Peg_V_Ha}a),
equivalent widths measured from H$\alpha$ (Fig.~\ref{V374Peg_V_Ha}b), and the 
$B$-$V$ (Fig.~\ref{V374Peg_V_Ha}c) and $V$-$I$ (Fig.~\ref{V374Peg_V_Ha}d) 
colours are shown. The $V$-band observations are plotted with different symbols
for the observations from different nights. Very rapid night-to-night 
variations are seen in the shape of the minimum around phases 0.3--0.6. Also, 
changes in the maximum around the phase 0.8 are seen during the observations 
spanning seven nights. As for EY~Dra, the $B$-$V$ colour of V374~Peg is 
basically flat and does not show marked variability. Between phases 0.9--0.3
the $B$-$V$ colour index shows high scatter because of the flares concentrating
in this part of the light-curve (see Fig.~\ref{flare2}). The $V$-$I$ on the 
other hand again clearly shows the locations of the main cool spots in the 
photosphere around phase 0.5. 

The comparison of the chromospheric and photospheric activity in V374~Peg does 
not show as clear a picture as in the case of EY~Dra. As can be seen from 
Fig.~\ref{V374Peg_V_Ha} a slight increase in the chromospheric emission is seen
at the phases of the photospheric spots. But, this increase is very weak, 
and the H$\alpha$ observations do not cover the phases of the main active 
region, thus making it impossible to draw clear conclusions on the possible 
correlation.

\subsection{Comparison of activity patterns in EY~Dra and V374~Peg}

EY~Dra and V374~Peg are both M dwarfs, but on different sides of the mass limit
for full convection. EY~Dra still has a small radiative core, whereas V374~Peg 
is fully convective. This change in the interior structure of a star plays an
important role in the dynamo operation. Thus, it is very interesting to 
compare the activity seen in these two stars. 

The spot configuration in EY~Dra shows two active regions, a configuration that
has been reported also in earlier papers (see, e.g., Korhonen et 
al.~\cite{Kor_EYDra}; Vida et al.~\cite{vida10}). The light-curve changes on 
long time scale (a 350~day long activity cycle has been reported by Vida et 
al.~\cite{vida10}), but is stable from night-to-night throughout more
than 80 days (i.e., $>$160 rotations) of our observations (see 
Fig.~\ref{EYDra_JD} and Fig.~\ref{EYDra_V}). In V374~Peg the overall
activity seems to stay stably at the same location for a long time (Morin et 
al.~\cite{morin08}), but the night-to-night variability shown here in the 
photometric observations from six nights is large. This implies that in 
V374~Peg fast rearrangement of spots occurs, but the active longitude 
itself stays at the same location on the stellar surface.

In EY~Dra clear correlation between the photospheric spots, and chromospheric 
emission is seen: enhanced emission occurs at the phase of the spots. In 
V374~Peg hints of similar correlation are seen, but unfortunately there are 
not enough chromospheric data from the phases of photospheric spots to confirm 
this behaviour.

Both of the stars show flaring activity. During the 12 nights of observations 
presented here no flares were seen in EY~Dra, but several were reported in the 
long-term monitoring by Vida et al.~(\cite{vida10}). In V374~Peg the current 
data obtained during nine nights, shows two large flares: one seen in the 
photometric observations, and another in the H$\alpha$ line. In addition, 
twelve smaller flares are detected in the broadband photometry. This implies 
that flares in V374~Peg are numerous, and occur more frequently than in EY~Dra.

The spectropolarimetric observations of M dwarfs show different magnetic 
configurations for fully convective and not fully convective stars. Donati et 
al. (\cite{donati08}) show that early M dwarfs, which still have a small 
radiative core, show predominantly toroidal and non-axisymmetric poloidal 
configurations, and have strong surface differential rotation. In the sample of
Donati et al. (\cite{donati08}) only the lowest mass early M dwarf showed 
dominantly axisymmetric poloidal fields, as is seen for the mid-M dwarfs which 
have masses close to full convection limit (Morin et al. \cite{Morin_midM}). 
Also, the early M dwarfs show long-term variability (Donati et al. 
\cite{donati08}) whereas the mid-M dwarfs exhibit stable active region 
configurations (Morin et al. \cite{Morin_midM}). The observations presented 
here seem to agree with this picture. But the fully convective stars, which do 
not generally seem to show long-term changes in their spot configurations, can 
still show fast short-term changes as is seen in V374~Peg. This implies that 
even if the active region is stable for a long time, the exact spot 
configuration within the active region itself changes on short time scales.

\section{Conclusions}

From the broad band photometry and intermediate resolution spectroscopy in the 
H$\alpha$ region the following conclusions can be drawn:

\begin{itemize}
\item The low resolution spectrum shows that GSC~02038-00293 is a K-type
  star. The obtained effective temperature of $4750\pm 250$~K, based on
  the intermediate resolution spectra, agree with the mid-K spectral type. 
\item GSC~02038-00293 has $v\,\sin i = 90\pm10$\kms.
\item Time variability is observed in the strength of the H$\alpha$ line in 
  GSC~02038-00293, indicating that this star is chromospherically active.
\item Active regions on EY~Dra are centred at phases 0.48 and 0.80 in 2008
  June--August.
\item Photospheric spots in V374~Peg occur at phases 0.3--0.6. 
\item Very strong night-to-night variation in the exact shape of the 
  light-curve minimum is seen in V374~Peg. This behaviour implies that even if 
  the activity is stably located at certain region of the star the exact spot 
  configuration within the active region changes rapidly.  
\item Two large flares, one detected in the broadband photometry and one in 
  H$\alpha$ observations, are seen on V374~Peg during the nine nights the data 
  presented here cover. In addition twelve smaller flares are seen in the 
  photometric observations.
\item The enhanced chromospheric emission in EY~Dra occurs at the same phases 
  as the photospheric spots, indicating plages collocated with the starspots, 
  as is also observed in the Sun. V374~Peg also shows some hints of this 
  behaviour but from the data presented here no firm conclusions can be drawn 
  for this issue.  
\end{itemize}

\acknowledgements
The spectroscopic observations used in this work were obtained during the 7th 
NEON observing school which was organised on La Palma, Spain, 2008 June 23 -- 
July 05. The school was essentially funded by a Marie Curie grant from the
European Commission under the FP7 program (Grant MSCF-CT-2004-012701), with 
complementary support from Opticon (Optical Infrared Coordination Network), an 
I3 Network under FP7 also, and moral support from the European Astronomical 
society. The authors would like to thank Prof. Michel Dennefeld, Coordinator of
the NEON schools, for organising this school. The financial support of OTKA 
grant K-81421 is acknowledged. KV was supported by the E\"otv\"os Scholarship 
of the Hungarian State. SM is supported by ORSAS, UK and the University of 
Birmingham, UK. SM also acknowledges the financial support provided by NEON. 
Nordic Optical Telescope is operated on the island of La Palma jointly by 
Denmark, Finland, Iceland, Norway, and Sweden, in the Spanish Observatorio del 
Roque de los Muchachos of the Instituto de Astrofisica de Canarias. ALFOSC is 
owned by the Instituto de Astrofisica de Andalucia (IAA) and operated at the 
Nordic Optical Telescope under agreement between IAA and the NBIfAFG of the 
Astronomical Observatory of Copenhagen.



\begin{thebibliography}{}

\bibitem[1978]{aln}
  Al-Naimyi, H.M.: 1978, ApSS, 53, 181
	
\bibitem[1961]{bab} 
  Babcock, H.W.: 1961, ApJ, 133, 572

\bibitem[2001]{bar_col} 
  Barnes, J.R., Collier Cameron, A.: 2001, MNRAS 326, 950 

\bibitem[2005]{barnes05} 
  Barnes, J.R., Cameron, A.C., Donati, J.-F., James, D.J., Marsden, S.C.,
  Petit, P.: 2005, MNRAS 357, L1 

\bibitem[2001]{bat_ibr} 
  Batyrshinova, V.M., Ibragimov, M.A.: 2001, Astronomy Letters 27, 29
	
\bibitem[2006]{ber_fra}	
  Bernhard, K., Frank, P.: 2006, IBVS 5719

\bibitem[2009]{biazzo09}
  Biazzo, K., Frasca, A., Marilli, E., Covino, E., Alcal{\~a}, J. M., 
  {\'C}akirli, {\"O}., Klutsch, A., Meyer, M.R.: 2009, A\&A 499, 579

\bibitem[1997]{cha_bar} 
  Chabrier, G., Baraffe, I.: 1997, A\&A 327, 1039 

\bibitem[2006]{cha_kue} 
  Chabrier, G., K\"uker, M.: 2006, A\&A 446, 1027 

\bibitem[1998]{del_etal98}
  Delfosse, X., Forveille, T., Perrier, C., Mayor, M.: 1998, A\&A 331, 581 

\bibitem[2000]{del_etal00} 
  Delfosse, X., Forveille, T., S{\'e}gransan, D., Beuzit, J.-L., Udry, S., 
  Perrier, C., Mayor, M.: 2000, A\&A 364, 217 

\bibitem[2006]{dobler06} 
  Dobler, W., Stix, M., Brandenburg, A.: 2006, ApJ 638, 336 

\bibitem[2006]{donati06} 
  Donati, J.-F., Forveille, T., Cameron, A.~C., et al.: 2006, Science 311, 633

\bibitem[2008]{donati08}
  Donati, J.-F., Morin, J., Petit, P., et al.: 2008, MNRAS 390, 545
	
\bibitem[2007]{drr}
  Dragomir, D., Roy, P., Rutledge, R.~E.: 2007, AJ 133, 2495

\bibitem[1998]{eibe}
  Eibe, M.T.: 1998, A\&A 337, 757

\bibitem[2008]{CABS08}
  Eker, Z., Ak, N. Filiz, Bilir, S., et al.: 2008, MNRAS 389, 1722

\bibitem[2005]{els_kor} 
  Elstner, D., Korhonen, H.: 2005, AN 326, 278 

\bibitem[2007]{fra_ber}	
  Frank, P., Bernhard, K.: 2007, Open European Journal on Variable Stars, vol. 
  0071, Issue 1, p.1
	
\bibitem[2008a]{frasca_bin}
  Frasca, A., Biazzo, K., Ta\c{c}, G., Evren, S., Lanzafame, A.C.: 2008, A\&A 
  479, 557

\bibitem[2008b]{frasca_LQHya}
  Frasca, A., K{\H o}v{\'a}ri, Zs., Strassmeier, K.G., Biazzo, K.: 2008, A\&A 
  481, 229

\bibitem[1998]{gre_rob} 
  Greimel, R., Robb, R.M.: 1998, IBVS 4652, 1

\bibitem[1994]{SPECTRUM}
  Gray, R.O., Corbally, C.J.: 1994, AJ 107, 742

\bibitem[1994]{jef94} 
  Jeffries, R.D., James, D.J., \& Bromage, G.E.: 1994, MNRAS 271, 476

\bibitem[2007]{Kor_EYDra}
  Korhonen, H., Brogaard, K., Holhjem, K., Ramstedt, S., Rantala, J., 
  Th{\"o}ne, C.C., Vida, K.: 2007, AN 328, 897

\bibitem[2007]{kovari07}
  K\H{o}v{\'a}ri, Z., Vilardell, F., Ribas, I., Vida, K., van Driel-Gesztelyi, 
  L., Jordi, C., Ol{\'a}h, K.: 2007, AN 328, 904 

\bibitem[2005]{kue_rue} 
  K{\"u}ker, M., R\"udiger, G.: 2005, AN 326, 265 

\bibitem[1993]{kurucz}
  Kurucz, R.L.: 1993, Kurucz CD No. 13

\bibitem[1969]{lei} 
  Leighton, R.B.: 1969, ApJ 156, 1

\bibitem[2008a]{morin08}
  Morin, J., Donati, J.-F., Forveille, T., et al.: 2008, MNRAS 384, 77

\bibitem[2008b]{Morin_midM}
  Morin, J., Donati, J.-F., Petit, P., et al.: 2008, MNRAS 390, 567

\bibitem[2007]{WASP}	
  Norton, A.J., Wheatley, P.J., West, R.G., et al.: 2007, A\&A 467, 785

\bibitem[2006]{olah06}
  Ol{\'a}h, K., Korhonen, H., K\H{o}v{\'a}ri, Zs., Forg{\'a}cs-Dajka, E.,
  Strassmeier, K.G.: 2006, A\&A 452, 303

\bibitem[1955]{parker} 
  Parker, E.~N.: 1955, ApJ 122, 293

\bibitem[1995]{robb_card}
  Robb, R.M., Cardinal, R.D.: 1995
  IBVS, 4270

\bibitem[2008]{sav_str}
  Savanov, I.S., \& Strassmeier, K.G.: 2008, AN 329, 364

\bibitem[2008]{skelly}
  Skelly, M.B., Unruh, Y.C., Cameron, A. Collier, Barnes, J.R., Donati, J.-F., 
  Lawson, W.A., Carter, B.D.: 2008, MNRAS 385, 708

\bibitem[2007]{vida07} 
  Vida, K.: 2007, AN 328, 817

\bibitem[2010]{vida10} 
  Vida, K., Ol{\'a}h, K., K{\H o}v{\'a}ri, et al.: 2010, AN 331, 250

\end{thebibliography}
\end{document}